# Making Quantum Key Distribution a Commodity: The All-Silicon Approach

Bernhard Schrenk
*Center for Digital Safety and Security*
*AIT Austrian Institute of Technology*
Vienna, Austria
bernhard.schrenk@ait.ac.at

*Abstract:* **The dawning of the quantum age makes quantum key distribution (QKD) an indispensable necessity for our global communication infrastructure. The realization of an all-silicon QKD transmitter supplied by a light source that is native to silicon integration platforms is seen as a disruptive step towards the pervasive introduction of QKD in new applications, which up to now have not been explored due to the missing credentials of quantum optics as a cost-effective and highly miniaturized technology.**

## I. Introduction

There is no room for mishap in the realm of communication networks, and thus the highest degree of security should be enjoyed. Burgeoning quantum computing technology has already picked up the wave of systematic and continuing developments in integrated circuit technology, which is now propelling its rapid advancement. QKD is a response to this looming threat of breaking asymmetrical encryption. It provides an information-theoretic secure (ITS) key generation method for which the laws of nature rather than computational functions provide a secure key between two parties. However, security does not stop at the perimeter of metro and core networks. As we approach short-reach networks and cost-sensitive segments, non-ITS technologies such as post-quantum cryptography or physical-layer security replace QKD for the mere fact that quantum-optic systems do not yet offer credentials as a low-cost element that can be seamlessly integrated as a pervasive ITS solution. Unlocking mass QKD deployment in these "commodity" domains is currently stalled due to two primary reasons:

First, QKD demonstrations have aimed at an extension of performances in terms of secure-key rate (SKR) and reach. This motivation is inspired by first pilot QKD deployments that now furnish optical long-haul and metro networks – two domains where neither cost efficiency nor an ultra-small form factor are a pressing requirement. However, performances such as a 1-Mb/s SKR might exceed the demands by mass applications by far: Owing to the high classical data rates that place one-time pad encryption out of question, QKD system integration instead considers a fast AES key renewal for securing data traffic. Under the so-called NIST limit, for which one new 256-bit AES key is required for every 64-GB chunk of data, a SKR of 1 Mb/s would be able to secure a classical link capacity of >250 Tb/s. This is exceeding the requirements for commodity QKD by far as it would equate, for example, to more than 10000 wired 50-Gb/s access connections.

Second, research has responded to miniaturization of QKD systems through a wide range of integration platforms [1-7], each of them being characterized by unique advantages – yet following a purely photonics-oriented approach. To date, there is no demonstration that would feature a monolithic integration scheme compatible with micro-electronic co-integration to gracefully blend quantum-optics with classical digital systems. Even though monolithic InP integration has been demonstrated [5, 6], the QKD sub-system would reside as a chiplet that is co-packaged with a silicon electronic application-specific integrated circuit (ASIC). Other works, which leverage the silicon ecosystem for photonic integration, require an external optical power supply or again, a hybrid integration with InP for this purpose. This implies complex die-level assembly with precise waveguide-to-waveguide alignment and additional wiring, together with hermitic packaging requirements due to the use of III-V semiconductors. None of these approaches is competitive enough for low-cost commodity applications.

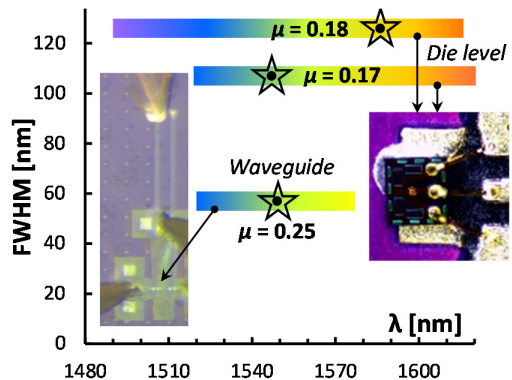

Fig. 1. Emission of die-level and waveguide-based silicon light sources.

## II. All-Silicon Quantum Optics as a Commodity

A possible path towards cost-efficient and highly miniaturized QKD is an all-silicon approach for the constituent quantum-optic circuits. Silicon – though being abundant as a material and the prime choice of the chip industry – does not offer light generation due to its indirect semiconductor bandgap, which prevents a native optical gain medium. This spoils a monolithic photonic-electronic integration scheme since the optically passive nature of silicon necessitates a complex and costly hetero-integration of III-V semiconductor dies (e.g., InP) or the hybrid integration of silicon with III-V chiplets. In the quantum realm, however, the required optical power levels are 8-9 orders-of-magnitudes lower than used for classical light sources in optical telecommunications or photonic sensing.

Figure 1 shows very recent accomplishments [8, 9] on silicon light emitters. These take advantage of a quasi-direct bandgap structure in silicon photonic waveguide platforms to attain a compact light source with an emission strong enough to optically power a QKD circuit and include (1) a die-level top-emitting SiGe pin structure and (2) a waveguide SiGe pin structure coupled to a 1D grating coupler on a silicon-on-insulator platform. The die-level sources are characterized by a more than 100-nm wide C+L band emission centered at 1586 and 1547 nm, respectively. The waveguide-based source shows a narrower, 57-nm wide spectrum at 1549 nm. Though these spectra are very wide due to the omission of spectral tailoring elements situated at the pin junction, they align well with mature WDM technology of the C-and L-bands. Moreover, the power delivered by the die- and waveguide-based devices are 23 and 34 pW, respectively. This equates to a mean photon number of $\mu = 0.17$ to $0.25$ photons/symbol, if a symbol rate of $R_{sym} = 1$ GHz is considered for the quantum transmitter. These emission levels need to be brought in context with the power budget of a QKD transmitter, as discussed in Table 1. Originating from the corresponding single-photon emission levels and given a 1-GHz symbol rate and Poissonian photon statistics (for which a set $\mu$-value of 0.1 photons/symbol ensures that there is no more than one photon in a transmitted pulse), the nominal launch power for the QKD transmitter would be -79 dBm. The means that the silicon light source provides a headroom of 2.5 dB (die-level) and 4 dB (waveguide-based) to accommodate for the loss of the modulator serving the quantum state preparation.

| Power budget | Source | ❶ Die-level | ❷ Waveguide | Unit |
|---|---|---|---|---|
| QKD TX | Single photon: λ | -159.0 | -158.9 | dBm |
| QKD TX | TX rate: 1 GHz | 90 | 90 | dB |
| QKD TX | Poisson: µ = 0.1 | -10 | -10 | dB |
| QKD TX | TX launch | -79.0 | -78.9 | dBm |
| Native Source | Emitted power | -76.5 | -74.9 | dBm |
| Native Source | Equivalent µ | 0.18 | 0.25 | |
| Native Source | Margin to µ = 0.1 | 2.5 | 4.0 | dB |
| Filtered Source | Filtering loss: 2 nm | 16.6 | 14.7 | dB |
| Filtered Source | Effective µ | 0.0039 | 0.0088 | |
| Filtered Source | Margin to µ = 0.1 | -14.1 | -10.7 | dB |

Table 1. Power budget for silicon-sourced QKD.

## III. QKD with Silicon Optical Power Supply

The incoherent nature of light emission restricts the protocol that can be chosen to polarization-encoded BB84, since no phase encoding can be applied. However, depolarization effects along optical fibers will inevitably lead to a degradation of the quantum signal [10]. The light emission of the silicon source needs to be filtered to narrow its spectrum and mitigate depolarization – at the expense of filtering losses. For light with a bandwidth of $\delta\lambda = 2$ nm, these losses amount to 16.6 and 14.7 dB, respectively. As a consequence, the power margin to the nominal transmit level of $\mu = 0.1$ photons/symbol falls to -14.1 and -10.7 dB, meaning that the light output of the silicon light source does not yet allow to exploit the same capacity in terms of SKR as it would apply to III-V laser-sourced QKD.

Still, as Fig. 2 shows, a secret key can be established over short-reach networks. For the die-level source followed by I/Q based polarization encoding [10], operation below the 11% QBER threshold can be facilitated for spectral linewidths of up to 2 nm (●). Moreover, a secure-key rate of 0.37 kb/s/detector has been obtained over 1 km using the waveguide source (△), which would allow to secure a classical channel capacity of 745 Gb/s under the NIST limit. Operation over longer link reach would require co-integrated spectral tailoring features for the silicon source to alleviate its output from passive filtering losses.

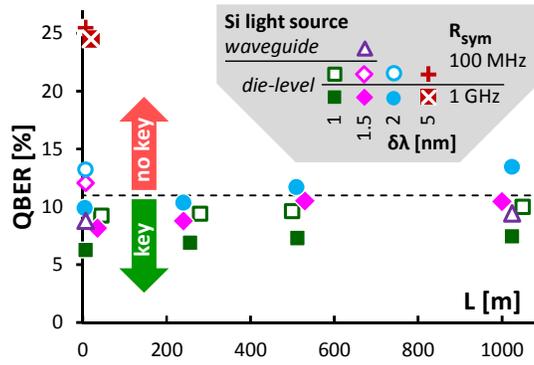
Fig. 2. Obtained QBER for QKD powered through silicon light sources.

## IV. Conclusion

The provision of an all-silicon solution for QKD through monolithic integration of the light source paves the way for a vast simplification of the overall QKD system, as it enables us to strip off complexity due to an external III-V based power supply. At the same time, it mitigates costly waveguide-to-waveguide assembly and hermetic packaging inherent to hybrid III-V/Si integration. The provision of a silicon light source can enable QKD to invigorate ICT segments as a pervasive ITS solution, especially in domains where non-ITS primitives still prevail, such as intra-facility or enterprise networks, industry environments, the public infrastructure, and IoT or personal wireless devices empowered through free-space optical key exchange.

**Acknowledgement:**

This work has received funding from the EU's H2020 research and innovation programme under the GA No 101017733 and by the Austrian research promotion agency FFG through the QISS·ME project (grant no. FO999906040).